\documentclass[twocolumn,superscriptaddress,nofootinbib,floatfix,aps,prb,citeautoscript,longbibliography]{revtex4-2}
\usepackage[utf8]{inputenc}
\usepackage[T1]{fontenc}
\usepackage{lineno}

\usepackage{hyperref}
\usepackage{graphicx}
\usepackage{booktabs}
\usepackage{color}
\usepackage{array}
\usepackage{dcolumn}
\usepackage{bm}
\usepackage{multirow}
\usepackage[version=4]{mhchem}
\usepackage{cleveref}

\usepackage[breakable, theorems, skins]{tcolorbox}

\begin{document}

\newcommand{\tb}[1]{\textbf{#1}}

\newcommand{\bochum}{Research Center Future Energy Materials and Systems of the University Alliance Ruhr and Interdisciplinary Centre for Advanced Materials Simulation, Ruhr University Bochum, Universitätsstraße 150, D-44801 Bochum, Germany}

\newcommand{\coimbra}{CFisUC, Department of Physics, University of Coimbra, Rua Larga, 3004-516 Coimbra, Portugal}

\author{Tiago F. T. Cerqueira}
\affiliation{\coimbra}
\author{Haichen Wang}
\author{Silvana Botti}
\email{silvana.botti@rub.de}
\author{Miguel A. L. Marques} 
\email{miguel.marques@rub.de}
\affiliation{\bochum} 

\date{\today}

\title{A non-orthogonal representation for materials based on chemical similarity}

\begin{abstract}
We present a novel approach to generate a fingerprint for crystalline materials that balances efficiency for machine processing and human interpretability, allowing its application in both machine learning inference and understanding of structure-property relationships. Our proposed material encoding has two components: one representing the crystal structure and the other characterizing the chemical composition, that we call Pettifor embedding. For the latter we construct a non-orthogonal space where each axis represents a chemical element and where the angle between the axes quantifies a measure of the similarity between them. The chemical composition is then defined by the point on the unit sphere in this non-orthogonal space. We show that the Pettifor embeddings systematically outperform other commonly used elemental embeddings in compositional machine learning models. Using the Pettifor embeddings to define a distance metric and applying dimension reduction techniques, we construct a two-dimensional global map of the space of thermodynamically stable crystalline compounds. Despite their simplicity, such maps succeed in providing a physical separation of material classes according to basic physical properties.
\end{abstract}

\maketitle

\section{Introduction}

The last decade has seen a remarkable surge in computational materials science, largely enabled by advances in high-throughput density functional theory and machine learning techniques~\cite{10.1038/s41524-019-0221-0,10.1088/2516-1075/ac572f}. These have greatly increased our knowledge and understanding of materials and contributed to the discovery of many compounds with improved properties. Inorganic materials databases catalog most experimentally verified compounds to date and millions of other hypothetical phases~\cite{alexandria,materialsproject,aflowlib,oqmdb,jarvis}. They also combine structural information with data on thermodynamic stability and a wealth of other mechanical, electronic, magnetic, etc. properties.

This explosion of data has brought with it new challenges. One particular challenge we address here is the digital representation of a compound that is both humanly understandable and suitable for material informatics techniques. In particular, these representations should allow the visualisation of material properties across different compositions and structural types, greatly extending the concept of structure maps~\cite{villars1989environment, pettifor1984chemical}. These were originally developed to find correlations between the structural type of a compound and the electron configuration of its constituents, providing insight into the structure of new or hypothetical compounds.  

For binary compounds (or multinary compounds where only two chemical species vary) it is often easy to produce two-dimensional maps where the $x-$ and $y-$ axes run through the periodic table (see, e.g., Refs.~\cite{10.1038/s41524-022-00868-7,10.1039/D0MA00999G,10.1021/acs.chemmater.2c01390,10.1038/s41467-023-38423-7,10.1038/s41467-023-38423-7}).  The order of the elements may simply reflect the atomic number, or some other ordering such as electronegativity or the Pettifor scale and its generalizations~\cite{pettifor1984chemical,pettifor1986structures,glawe2016optimal,10.1021/acs.jpcc.0c07857,10.1021/acs.chemmater.2c01390}. The latter is particularly interesting because it reflects the similarity between chemical elements and gives the systematic variation of the property over the binary composition range. Unfortunately, for ternary or multinary compounds, or when the dataset contains materials with different crystal structures, the production of such material maps is much more complicated.

In fact, it is very difficult to represent material space in a way that allows visualisation --- and interpretation --- of how material properties vary across a given set of compounds.  A common solution is to use machine learning embeddings, which are readily available from most neural network architectures. For example, in crystal graph networks~\cite{cgcnn} we can use the feature vector obtained after pooling the graph. These can then be used in conjunction with dimension reduction techniques~\cite{10.1080/14786440109462720,10.48550/arXiv.1802.03426} to produce two-dimensional (or higher dimensional) maps that can be used to visualise material properties over material space~\cite{cgcnn_view,crabnet_view}. Unfortunately, trained embeddings already contain information about the target property (or properties), which often complicates the interpretation of structure-property relationships.

To solve this problem, we need a representation of a material in a vector space (also sometimes called fingerprint, descriptor, or embedding) based only on its chemical composition and crystal structure, such that the distances between points (i.e. compounds) reflect the degree of similarity between these compounds. 

We divide the fingerprint into two parts. The first should be a representation of the crystal structure of the materials. Fortunately, there are already several structural fingerprints available in the literature. We will use the one available in \textsc{pymatgen}~\cite{pymatgen}, which measures the similarity between two structures based on local coordination information from all sites in the crystal structures~\cite{zimmermann2017assessing}. This representation meets our twin requirements of being machine-friendly and based on a solid human understanding of the underlying physics.

For the description of the chemical composition a common solution is to use a one-hot vector, where each element represents a chemical element, a vector constructed from the properties of the elements~\cite{cgcnn, Ward2018}, or machine-learned representations~\cite{mat2vec}. Unsurprisingly, the latter are more efficient for machine learning~\cite{crabnet} but are opaque and not interpretable by humans. We formulate therefore a new scheme that retains the usefulness of these machine-learned approaches and, at the same time, is simple and fully human understandable. 

We start with a measure of the similarity between chemical elements. This could be, for example, the Euclidean distance between the initial (untrained) embeddings of the chemical elements, as provided by \textsc{matminer}~\cite{Ward2018}. Instead we introduce a different measure based on the similarity scale proposed by some of us in Ref.~\cite{wang2021predicting}, which was constructed by simple statistical arguments using the experimentally known inorganic compounds present in ICSD~\cite{10.1107/S160057671900997X}. Using this similarity measure, we establish a non-orthogonal space where each axis corresponds to a chemical element and the angle between axes quantifies the similarity between elements. In this framework, chemical compositions are represented as points on the unit sphere of this non-orthogonal space. The details of the method to define our similarity metric and the corresponding composition embeddings, that we call Pettifor embeddings, are contained in \cref{sec:method}.

We remark that recent developments in representing chemical composition have explored alternatives to Euclidean distance metrics. Notable among these are approaches based on Earth Mover's Distance (EMD) for comparing compositions~\cite{hargreaves-ElMD-2020,zhang-GRID-2023}, which measure the minimal ``cost'' of transforming one composition into another, based on data-mined definitions of chemical similarity~\cite{glawe2016optimal,hautier-substitutions-2011}.
Our approach differs by maintaining an Euclidean distance, but in a non-orthogonal space where the angle between the axes, and therefore the distance between two compositions, is controlled by elemental similarities and emerges naturally from observed substitution patterns in experimental crystal structures. This approach captures richer information about elemental relationships than one-dimensional ordering schemes (like the modified Pettifor scale) used to define, e.g., the Element Mover's Distance (ElMD) ~\cite{hargreaves-ElMD-2020}, allowing us to distinguish between neighbors of the same order of a given element by their absolute distance.

Having defined the problem and outlined our approach, we now present results showing how our non-orthogonal representation captures meaningful chemical relationships, generates interpretable visualizations of material space, and provides powerful embeddings for machine learning applications, all while maintaining human-understandable connections to underlying chemical principles.

\section{Results}

\begin{figure}[tb]
    \centering
    \includegraphics[width=\columnwidth]{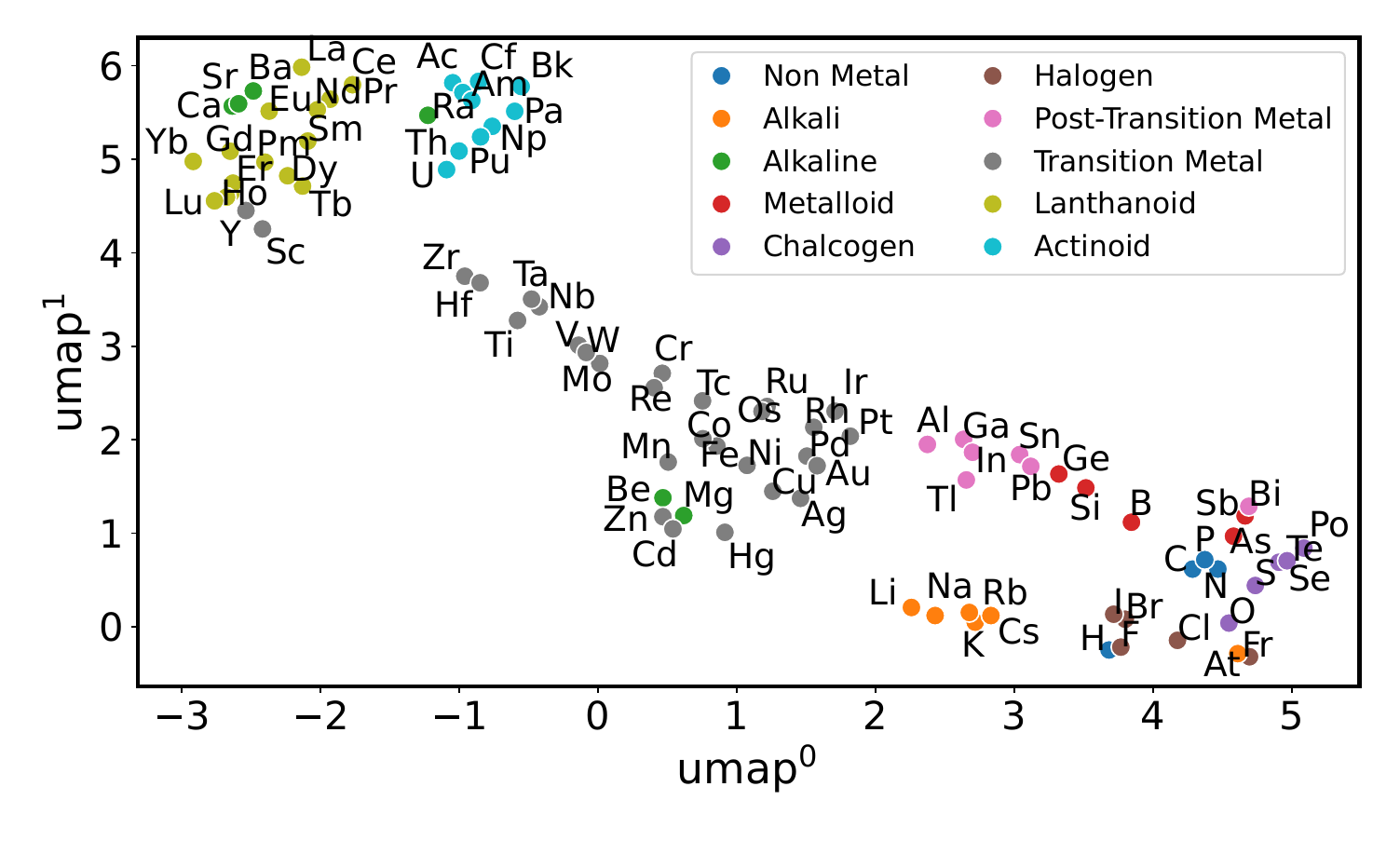} 
    \includegraphics[width=\columnwidth]{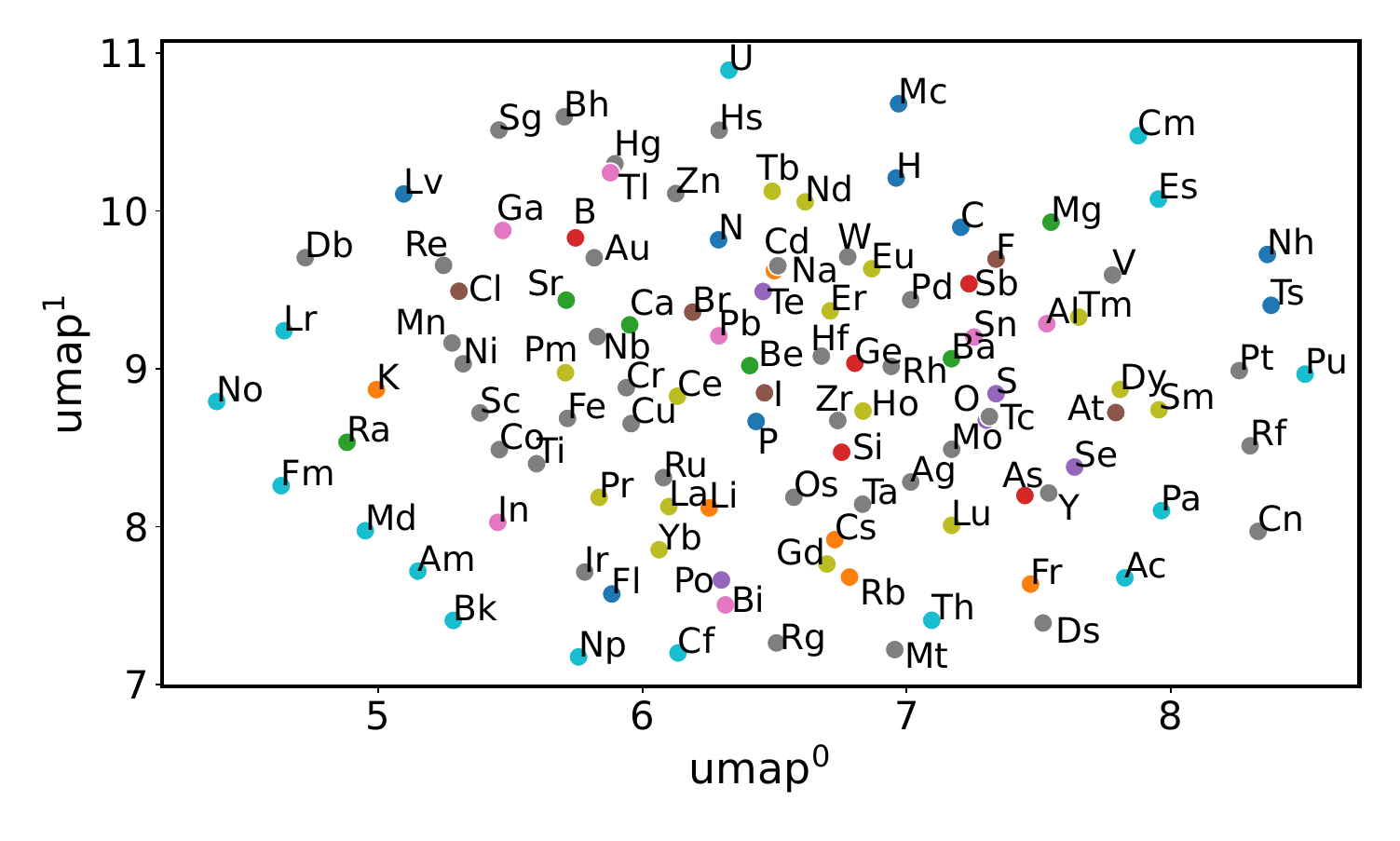} 
    \caption{Two-dimensional maps of the chemical elements obtained by reducing the dimensions of the Pettifor embeddings (top panel) and of the one-hot embeddings (bottom panel). The axes indicate the two dimensions returned by UMAP. The points are colored in both panels according to the group of the periodic table to which the elements belong. Comparison with other common embeddings is available in the ESI.}
    \label{fig:periodic_table}
\end{figure}

\begin{table*}[ht!]
    \centering
    \caption{Top three most similar elements to S, Si, and Ag based on ElMD~\cite{hargreaves-ElMD-2020} and C-GRID~\cite{zhang-GRID-2023}, as well as Euclidean distances using Mat2Vec~\cite{mat2vec}, Magpie~\cite{magpie}, CGCNN~\cite{cgcnn}, and our Pettifor embeddings. Distances are shown in parentheses. Note that the absolute values of distances can not be compared across different representations.}
    \begin{tabular*}{0.99\textwidth}{@{\extracolsep{\fill}} c|ccc|ccc|ccc}
    \toprule
        Metrics & \multicolumn{3}{c|}{S} & \multicolumn{3}{c|}{Si} & \multicolumn{3}{c}{Ag} \\
    \midrule
        ElMD & Se (1) & O (1) & Te (2) & Ge (1) & B (1) & Sn (2) & Cu (1) & Au (1) & Pd (2) \\
        C-GRID & Se (0.08) & Cr (0.10) & Be (0.12) & Ge (0.06) & As (0.17) & V (0.18) & Cu (0.07) & Au (0.28) & Li (0.30) \\
        Mat2Vec & N (3.0) & O (3.3) & C (3.4) & Ge (2.3) & Al (2.4) & Fe (2.9) & Au (2.2) & Cu (2.5) & Pd (3.1) \\ 
        Magpie & Ga (98.5) & P (98.6) & I (110.2) & Ni (59.4) & Co (98.1) & Fe (132.2) & Ge (54.2) & Pr (98.1) & La (105.4) \\
        CGCNN & Sb (1.73) & Lu (2.0) & Yb (2.24) & In (1.73) & Yb (2.24) & Hf (2.24) & Ce (1.0) & Yb (1.73)  & Hf (1.73) \\
        Pettifor & Se (0.77) & Po (0.88) & Te (0.94) & Ge (0.64) & Sn (0.96) & Ga (1.04) & Au (0.74) & Cu (0.85) & Pd (0.90)\\
    \bottomrule
    \end{tabular*}
    \label{tab:ele_distances}
\end{table*}

To demonstrate the effectiveness of our approach, we first computed the compositional fingerprints of the chemical elements, represented by the rows of the similarity matrix S\cal S S, as detailed in \cref{sec:method}. 

We can represent the periodic table by performing a dimension reduction of our compositional embeddings, using the Uniform Manifold Approximation and Projection (UMAP)~\cite{10.48550/arXiv.1802.03426}. The resulting two-dimensional map, which can be thought of as a data-mined and machine-learned periodic table, is shown in \cref{fig:periodic_table}. We have removed noble gases (which rarely form compounds) and have coloured the points according to the group of the periodic table. 

From the construction, we expect similar chemical elements to be close together in the map, although the reduction to only two dimensions may distort the distances in the $>80$-dimensional composition space. We also note that the actual shape of the map depends on the parameters of UMAP (such as the number of neighbours or the minimum distance used by UMAP to decide how closely points are packed together), but the relative distribution of chemical elements is quite robust. In the ESI it is possible to see the impact of different UMAP parameters on the clustering patterns and spatial distribution of chemical elements in the reduced feature space.

The top panel of \cref{fig:periodic_table} depicts the two-dimensional map of chemical elements obtained by reducing the dimensions of the Pettifor embeddings. This visualization clearly shows chemical elements with similar properties clustered together. The lanthanides and actinides appear in the top left region, transition metals occupy the middle area of the plot, while non-metals and alkali elements are predominantly positioned in the bottom right. This natural clustering emerges solely from the substitution patterns observed in crystal structures, without explicitly encoding traditional periodic table relationships.

The bottom panel of \cref{fig:periodic_table} presents, for comparison, a similar two-dimensional projection obtained using the one-hot~\cite{harris2015digital} encoding representation. Unlike our Pettifor embedding, which captures chemical similarities through angles between element vectors, the one-hot representation treats all elements as equidistant from each other in the original space. Other commonly used elemental embeddings, like Magpie~\cite{magpie}, Mat2Vec~\cite{mat2vec}, Jarvis~\cite{jarvis}, and CGCNN~\cite{cgcnn}, display an intermediate performance in comparison to the Pettifor and the one-hot embeddings, with Magpie achieving the second best representation. The corresponding two-dimensional maps are shown in Fig.~S1 of the ESI.

To further evaluate the effectiveness of our compositional embedding against existing approaches, we compared the elemental similarities (e.g., distances in the embedding space) captured by different methods. Table~\ref{tab:ele_distances} shows the top three most similar elements to S, Si, and Ag according to various metrics: ElMD~\cite{hargreaves-ElMD-2020}, the composition-only EMD used for the grouped representation of interatomic distances (GRID) of Ref.~\cite{zhang-GRID-2023}, here called for simplicity C-GRID, and Euclidean distances obtained from Mat2Vec\cite{mat2vec}, Magpie~\cite{magpie},  and our Pettifor embedding.

The metric derived from our elemental embeddings shows general agreement with ElMD, which is expected as both methods derive from the same chemical similarity matrix that was used to define the modified Pettifor scale~\cite{glawe2016optimal}.  However, significant differences emerge when compared with property-based embeddings like Magpie or text-mining approaches like Mat2Vec. For example, while our method identifies that selenium (Se) is chemically most similar to sulfur (S) with a distance of 0.77, Magpie unexpectedly suggests gallium (Ga) as most similar. The key advantage of our approach with respect to ElMD is its ability to capture nuanced differences in similarity. For instance, our embedding distinguishes between the distances of S to Se (0.77) and S to O (1.26), even if both Se and O are nearest neighbors of S in the 1D modified Pettifor scale, reflecting the much higher replaceability of S by Se than by O in real compounds. Furthermore, unlike embeddings derived from trained models, our representation remains independent of specific property prediction tasks, offering a more general framework for understanding chemical similarity.

\begin{figure}[tb!]
    \centering
    \includegraphics[width=\columnwidth]{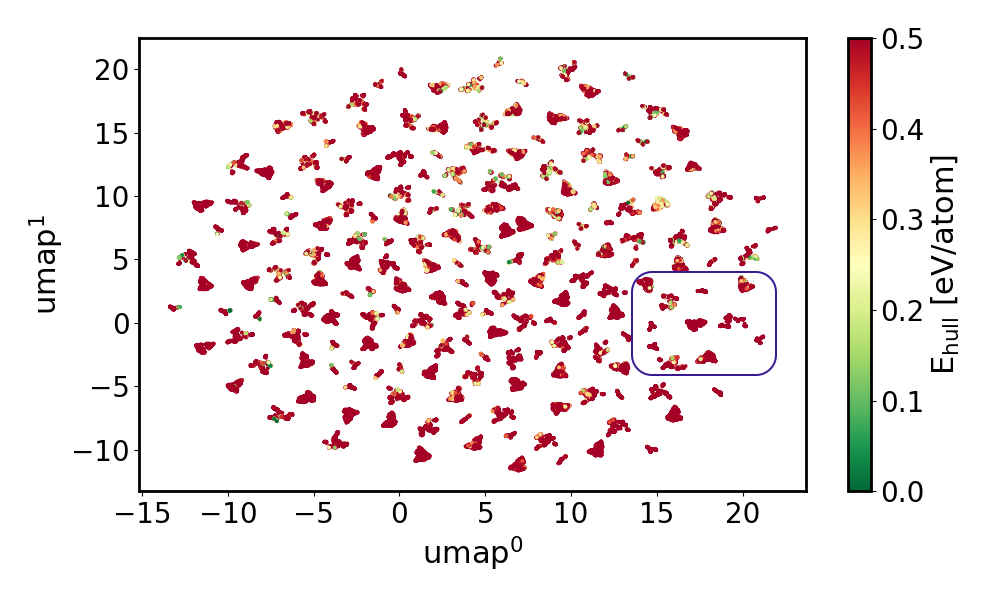}
    \includegraphics[width=\columnwidth]{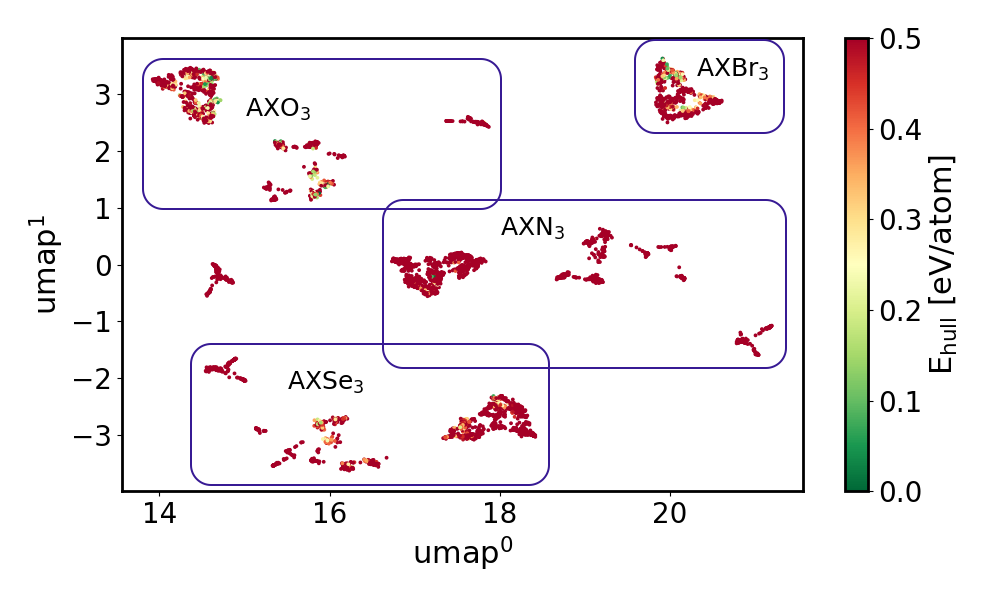}

    \caption{Top panel: Two-dimensional composition map of cubic \ce{ABC3} perovskites obtained by reducing the dimensions of the Pettifor embeddings with UMAP. For the plot we chose the fixed structure \ce{ABC3} or \ce{BAC3} with the lowest energy. Points are coloured according to the distance to the convex hull of the corresponding material, capped at 0.5~eV/atom. Bottom panel: Magnified view of the region containing AX\{O, N, Se, Br\}$_3$ perovskites, indicated by a blue line in the top panel. For each family, the main cluster contains compositions including transitions metals.}
    \label{fig:perovskites}
\end{figure}

As a first example of application of the Pettifor embeddings to crystalline materials, we show in \cref{fig:perovskites} two-dimensional maps of the dataset of perovskites with composition \ce{ABC3} from Ref.~\onlinecite{Schmidt2017}. We note that the compounds in this dataset all share the same structure, namely the cubic $Pm\bar{3}m$ (space group \#221), where the $1a$, $1b$, and $3c$ Wyckoff positions are respectively occupied by A, B, and C atoms, giving a primitive cell containing 5 atoms. The figure provides therefore only a map based on the chemical composition. As both compounds \ce{ABC3} and \ce{BAC3} have the same composition, and therefore the same Pettifor embeddings, we plot only the compound with the lowest energy. The points are coloured according to their distance from the convex hull of thermodynamic stability, showing stability trends within and across these perovskite subfamilies. Since the chemical composition is dominated by the atom C, the map is naturally divided into clusters, one for each C. The distribution of the clusters is then largely determined by the similarity between the C atoms. Within each cluster, there is a fine structure that reflects the similarity between the A and B atoms.

The bottom panel of \cref{fig:perovskites} provides a magnified view of the region containing AX\{O, N, Se, Br\}$_3$ perovskites (indicated by the blue line in the top panel). This zoomed view reveals distinct clustering patterns that are present within each perovskite family. For each cluster, we observe that compositions including transition metals form the main clusters, while other cations form smaller islands.
It is clear from the color code of \cref{fig:perovskites} that most cubic perovskites are highly unstable, but some clusters of higher stability can be seen. The well-known oxide perovskites \ce{ABO3} are centred at the coordinates $(15.3, 2.5)$. However, most stable (or near stable) compounds are inverted perovskites with an H, C, N, O, etc. in the Wyckoff 1b position. These can be seen as green dots usually at the boundaries of the clusters. 

By far the most stable systems are the inverted perovskites of the type \ce{ABCa3} around $(3.8, 18.8)$, \ce{ABSr3} around $(5.2, 15.9)$, \ce{ABSc3} around $(3.2, 8.8)$, etc. 

\begin{figure}[tb]
    \centering
    \includegraphics[width=\columnwidth]{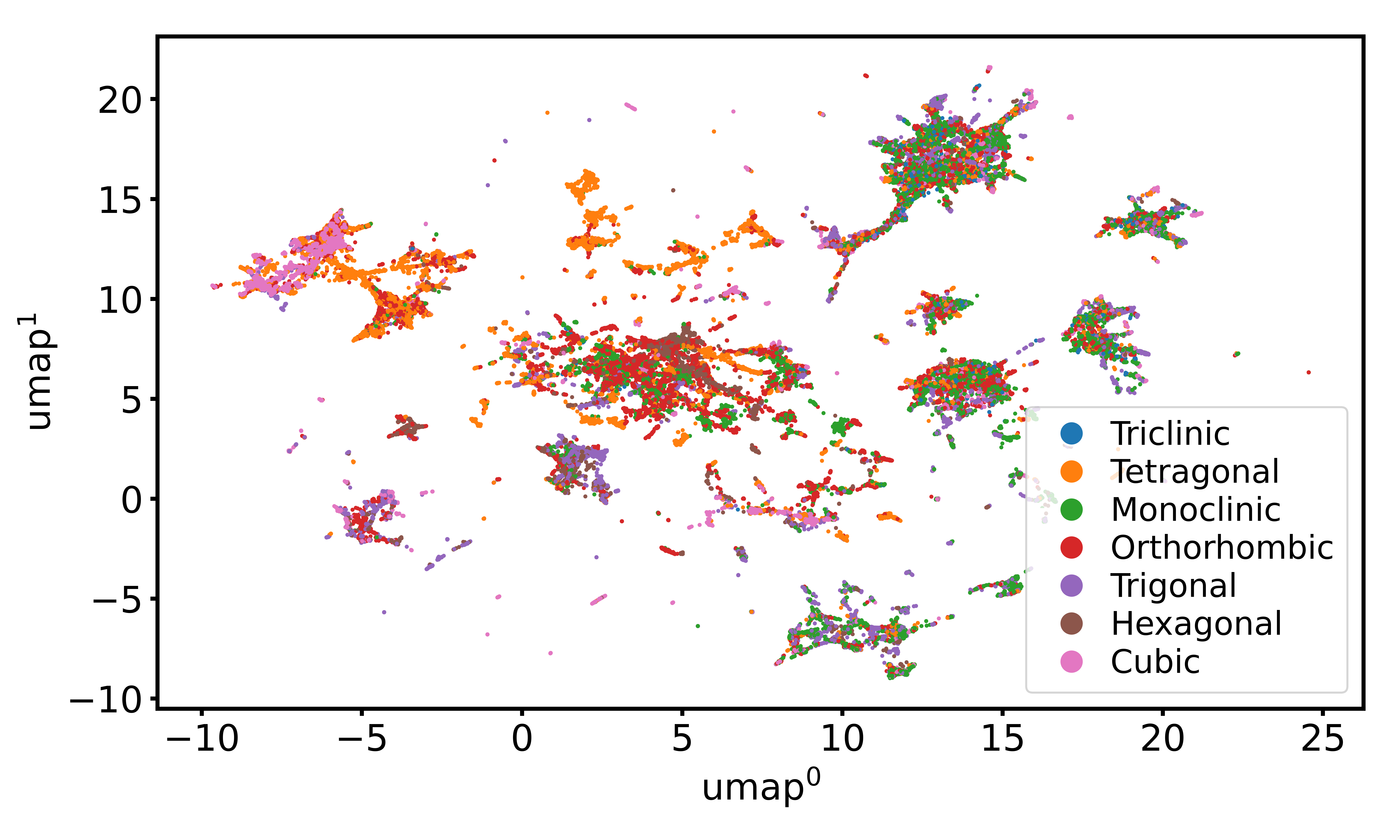}

    \includegraphics[width=\columnwidth]{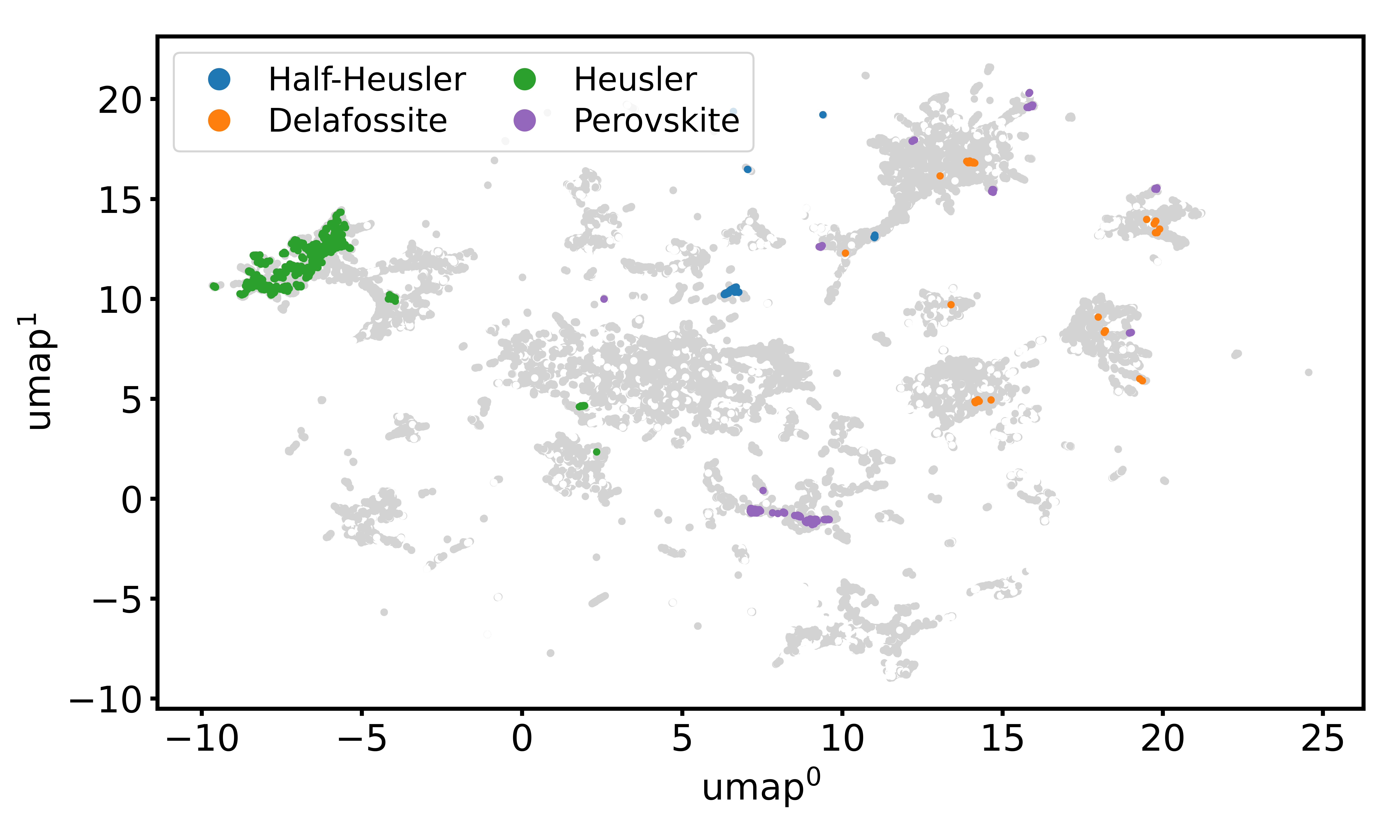}
    
    \caption{Materials on the convex hull of Alexandria, projected into two dimensions using UMAP on our Pettifor$\oplus$STR embeddings, taking into account both crystal structure and chemical composition. (Top) We distinguish by crystal system. (Bottom) We highlight some well-known materials families.}
    \label{fig:convex_hull}
\end{figure}

\begin{table*}[tbh]
\centering

\caption{Mean absolute error across various datasets~\cite{crabnet} comparing different feature representations for CrabNet: \textsc{mat2vec}, \textsc{magpie}, \textsc{cgcnn}, one-hot encoding\cite{harris2015digital}, and our proposed non-orthogonal representation. Note that for the classification tasks 'Exp is metal' and 'glass', the metric shown is Area Under the Receiver Operating Characteristic Curve (ROC AUC), where higher values indicate better performance. Detailed information on the benchmark datasets can be found in Ref.~\cite{crabnet}.}
\begin{tabular*}{0.95\textwidth}{@{\extracolsep{\fill}} lrlllll}
\toprule
Dataset & Set size & \textsc{mat2vec} & \textsc{magpie}  & \textsc{cgcnn} & \textsc{one-hot} & This work         \\

\midrule
steels\_yield                & 312    & 128        & 207       & 132        & \tb{126}    & 136            \\
jdft2d                       & 636    & 41.7       & 40.0      & \tb{37.8}   & 41.4       & 40.8            \\
phonons                      & 1265   & 78.3       & 134       & \tb{61.4}   & 82.3       & 73.2            \\
Aflow thermal expansion      & 3421   & 4.70$\times10^{-6}$ & 10.7$\times10^{-6}$ & $\mathbf{4.17\times10^{-6}}$ & 4.80$\times10^{-6}$ &  4.24$\times10^{-6}$         \\
Aflow thermal cond.          & 3422   & 2.50 & 3.30 & \tb{2.29} & 2.48 & 2.31              \\
Aflow bulk modulus           & 3428   & 9.89 & 21.4 & 9.95     & 9.97 & \tb{9.34}          \\
Aflow Debye temp.            & 3428   & 36.6 & 56.8 & \tb{33.7} & 37.0 & 34.2               \\
Aflow shear modulus          & 3428   & 10.0 & 15.4 & 9.62     & 10.0 & \tb{9.48}          \\
MP shear modulus             & 4328   & 13.2       & 15.9      & 12.4       & 13.2       & \tb{12.2  }          \\
MP bulk modulus              & 4414   & 12.5       & 21.8      & 11.7       & 12.6       & \tb{11.2  }          \\
MP elastic anisotropy        & 4431   & 8.29       & \tb{8.16}  & 8.16       & 8.16       & 8.23            \\
Exp E$_\mathrm{gap}$         & 4604   & 0.368      & 0.576     & \tb{0.335}  & 0.362      & 0.352         \\
dielectric                   & 4764   & 0.271      & 0.379     & 0.254      & 0.269      & \tb{0.254 }           \\
Exp is metal (\%)                 & 4921   & 95.6      & 93.2 & 96.6      & 95.7      & \tb{96.7}         \\
glass (\%)                       & 5680   & 90.4      & 69.9 & \tb{91.1}      & 90.7      & 90.1        \\
log$_{10}$($G_\mathrm{VRH}$) & 10987  & 0.105      & 0.138     & 0.0993     & 0.105      & \tb{0.0951}         \\
log$_{10}$($K_\mathrm{VRH}$) & 10987  & 0.0777     & 0.111     & 0.0756     & 0.0782     & \tb{0.0723}         \\
Aflow E$_\mathrm{gap}$       & 19330  & 0.331      & 0.451     & 0.321      & 0.330      & \tb{0.309 }           \\
Aflow energy/atom            & 19346  & \tb{0.109}  & 0.342     & 0.117      & 0.112      & 0.110            \\
MP E$\mathrm{hull}$          & 39663  & 0.0983     & 0.127     & 0.0960     & 0.0983     & \tb{0.0922}            \\
MP $\mu_b$                   & 39663  & 2.29       & 3.79      & 2.50       & 2.31       & \tb{2.22  }          \\
CritExam Ed                  & 59509  & 0.0651     & 0.0843    & 0.0630     & 0.0651     & \tb{0.0608}           \\
CritExam Ef                  & 59509  & \tb{0.0765} & 0.188     & 0.0821     & 0.0773     & 0.0800           \\
OQMD bandgap                 & 239125 & 0.0592     & 0.107     & 0.0593     & \tb{0.0580} & 0.0588         \\
OQMD energy/atom             & 239190 & \tb{0.0527} & 0.0696    & 0.0901     & 0.0525     & 0.0530         \\
OQMD form. Enthalpy          & 239190 & \tb{0.0423} & 0.0606    & 0.0530     & 0.0421     & 0.0426         \\
OQMD volume/atom             & 239190 & \tb{0.329}  & 0.421     & 0.405      & 0.329      & 0.330            \\
\bottomrule  
\end{tabular*}
\label{tab:crabnet}
\end{table*}  

We further benchmark the performance of our Pettifor embeddings to represent compounds in compositional neural network models. Specifically, we use the compositionally-restricted attention-based network (CrabNet~\cite{crabnet}), trained with different embeddings (\textsc{mat2vec}~\cite{mat2vec}, \textsc{magpie}~\cite{magpie}, \textsc{cgcnn}~\cite{cgcnn}, the one-hot representation, and Pettifor) for predicting a variety of materials properties. We present in Table ~\ref{tab:crabnet} results for 27 benchmarks~\cite{crabnet} covering a wide range of material properties and dataset sizes. We note that for the CGCNN embeddings, we used the initial atomic features and not the trained embeddings, which would change depending on the target training property. The results are averaged over 8 different runs in order to decrease the variability due to the training. We see from the table that our constructed representation yields on average the best results, while retaining simplicity and interpretability. Moreover, it performs particularly well for  small- and medium-sized datasets. This improved performance suggests that our representation effectively captures the essential chemical relationships relevant to materials properties, making it especially valuable for materials discovery problems where extensive training data may not be available. Of course, our fingerprint can also be used as a node embedding in graph neural networks (or in other architectures), but this is beyond the scope of this work.

To compute structure maps of compounds with arbitrary structures, we can concatenate the composition and structure fingerprints, obtaining the Pettifor$\oplus$STR embeddings. The two components can be combined with different weights, depending on whether one wants to give more importance to the composition or the structure. For simplicity, we have chosen equal weights for both descriptors in the following. 

\begin{figure}[tb]
    \centering

    \includegraphics[width=\columnwidth]{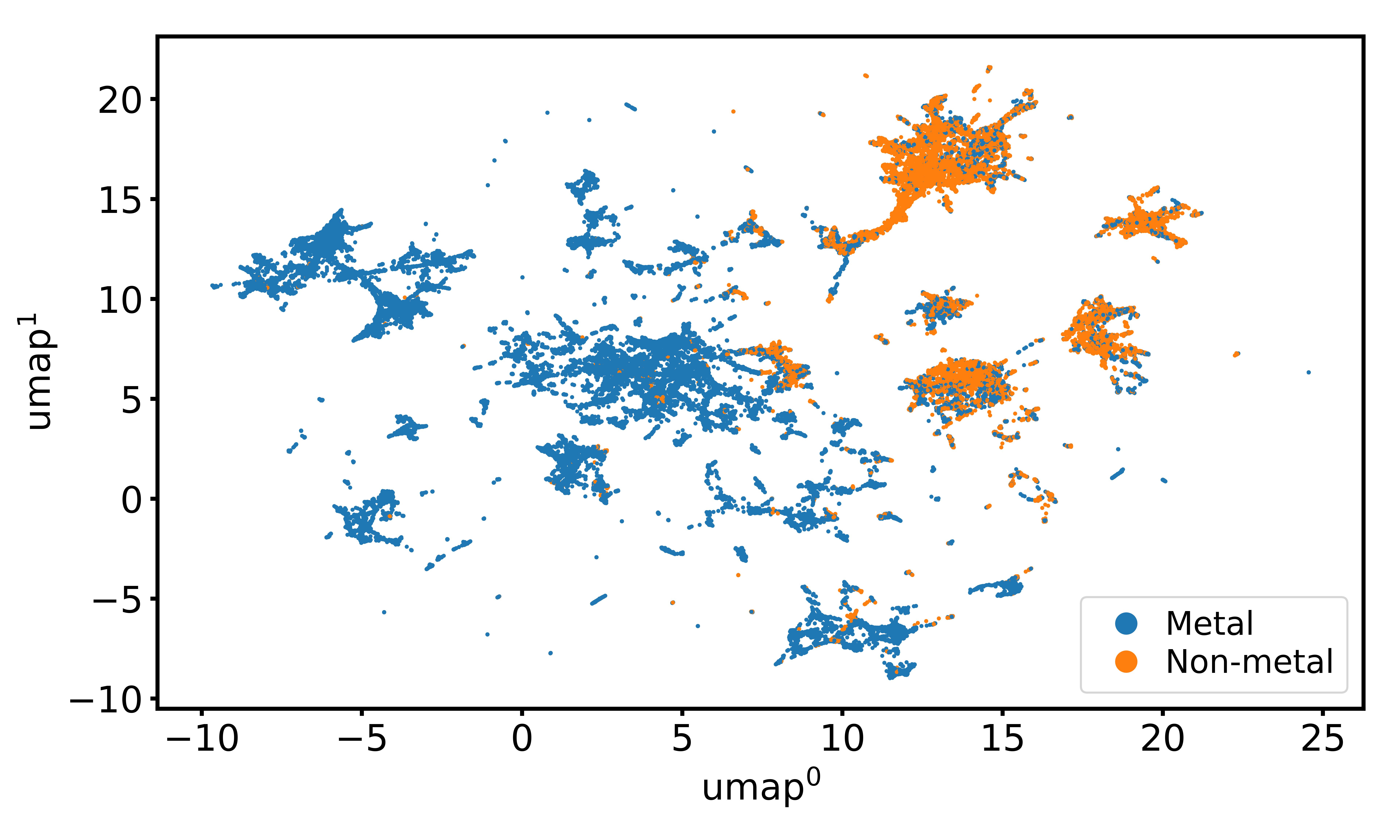}

    \includegraphics[width=\columnwidth]{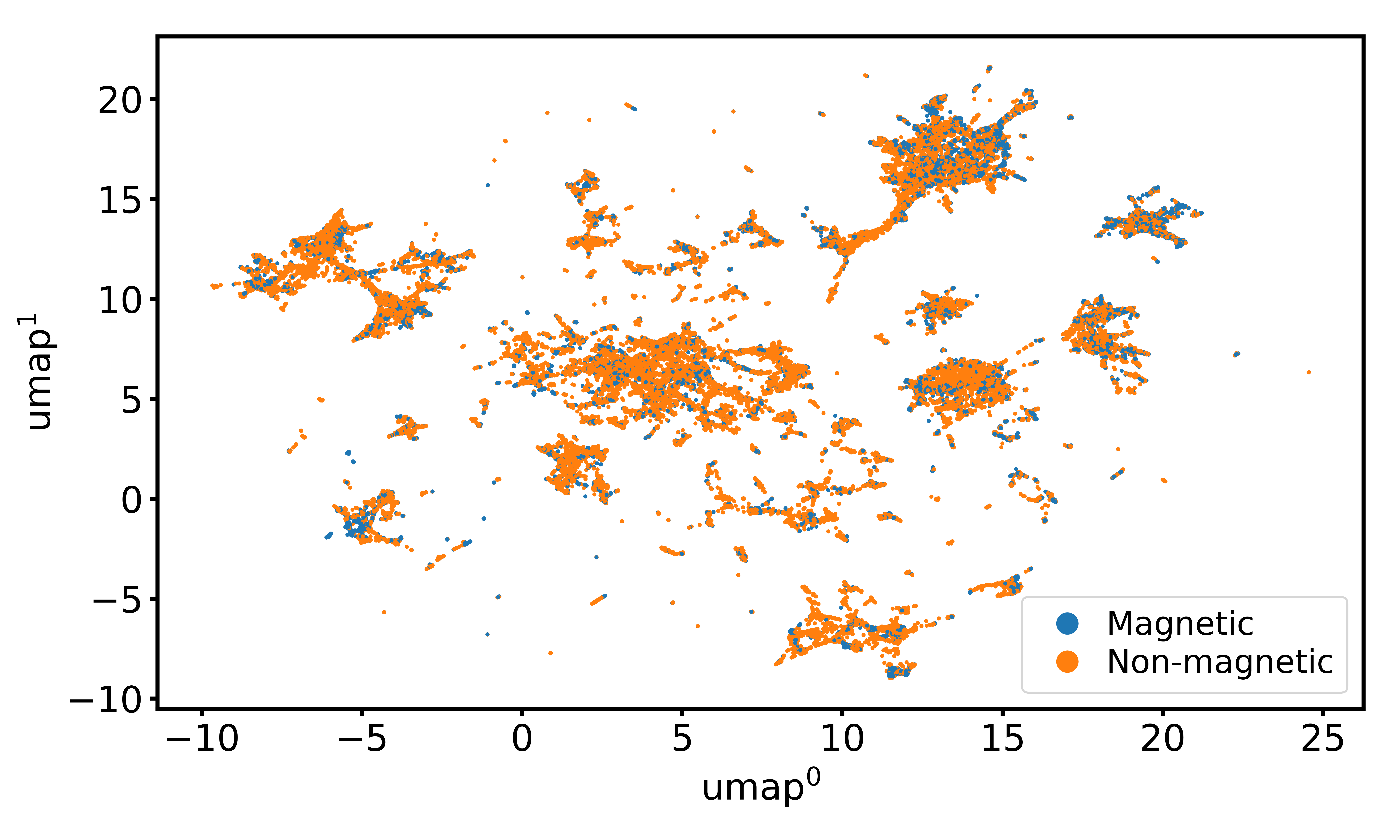}
    
    \caption{Materials on the convex hull of Alexandria, projected into two dimensions using UMAP with our Pettifor$\oplus$STR embeddings. We distinguish between metallic and non-metallic (top), as well as magnetic and non-magnetic materials (bottom).}
    \label{fig:convex_hull_2}
\end{figure}

As a first example of application of the Pettifor$\oplus$STR embeddings, we plot in \cref{fig:convex_hull} the map of all thermodynamically stable materials, i.e. compounds that lie on the convex hull of thermodynamic stability, found in the {\sc alexandria}~\cite{alexandria} database. This corresponds to over 115 thousand compounds with a wide variety of chemical compositions and structural types. To reduce the number of dimensions to two, we again use UMAP. We color the points according to the property that we want to highlight, e.g. the crystal system (\cref{fig:convex_hull}), the existence of an electronic band gap (top panel of \cref{fig:convex_hull_2}), or the existence of finite magnetic moments (bottom panel of \cref{fig:convex_hull_2}).

As a full discussion of the materials on the convex hull and their properties goes well beyond the scope of this work, we limit ourselves to a few general observations. UMAP divides the vast majority of the compounds into large ``continents'', with a few materials scattered in smaller islands. While the actual two-dimensional map is dependent on the parameters of UMAP, it is reasonable to interpret these smaller islands as more structurally and compositionally exotic compounds with few related materials on the convex hull.  It is interesting to see in \cref{fig:convex_hull} that, despite the extremely large diversity of the data set, the map provides a reasonable structural separation of compounds --- indeed, some of the continents and islands are dominated by compounds of a particular crystal system. Furthermore, other islands correctly identify similar structural motifs, such as the hexagonal and trigonal symmetry adopted by many layered materials, or the cubic, tetragonal, orthorhombic sequence often obtained for many compounds as symmetry decreases with, for example, decreasing temperature. 

UMAP combined with our fingerprint provides an excellent separation between metals and semiconducting or insulating compounds, as shown in the top panel of \cref{fig:convex_hull_2}. However, such a good separation is not visible in the bottom panel, which represents magnetic and non-magnetic compounds. This suggests that, unfortunately, this latter distinction will be much more difficult to predict with machine learning approaches than the previous properties.

An interesting question concerns the actual number of dimensions needed in practice to represent all the materials on the hull. Our Pettifor$\oplus$STR vector contains $61\times4=244$ elements to represent the structure, to which we add one extra dimension per chemical element. The question is how many of these $>300$ dimensions are really required. We can get this information by performing a principal component analysis of the data. We find that only 7 components are sufficient to represent 60\% of the variance in composition, while 6 are required to represent 60\% of the variance in composition and structure. If we ask for 95\%, we need 39 components for composition and 42 for composition and structure. This is much smaller than the original number of components, meaning that it may be possible to compress the information into a smaller feature vector that is more suitable for efficient machine learning of materials data. 

To demonstrate the practical utility of our Pettifor$\oplus$STR embeddings, we examine at last three concrete applications that illustrate how they can advance materials research and discovery.

\begin{figure}[tbh!]
\centering
\includegraphics[width=0.9\columnwidth]{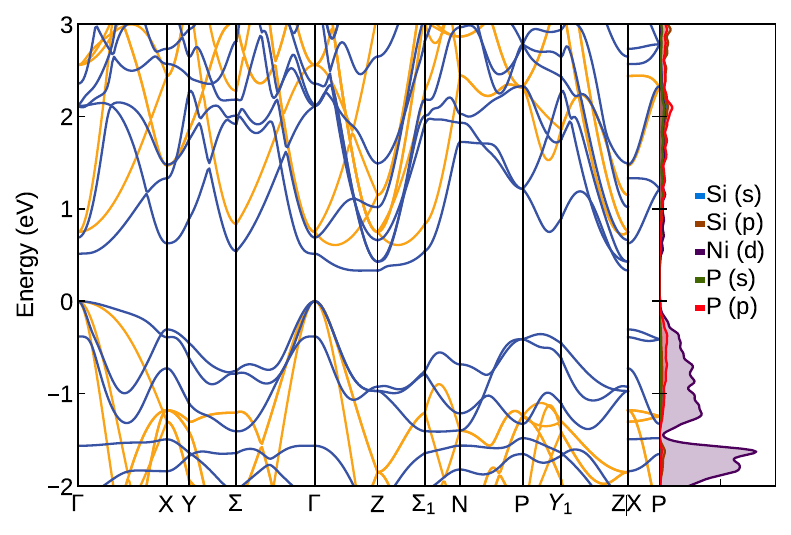}
\caption{Band structures of \ce{NiSi3P4} The upfolded band structure of Si is displayed in orange lines for comparison.}
\label{fig:similar_bands}
\end{figure}

First, we applied our embeddings to identify compounds with similar electronic properties. We consider as an example the crystalline materials that are closest to diamond silicon according to the distances in our embedding space. While Ge (distance 0.64) and SiC (distance 0.70) are predictably close to Si, our method also identifies less obvious compounds like \ce{NiSi3P4} (distance 0.71) that retain similar electronic properties. In \cref{fig:similar_bands} we show the band structure of the latter material. Figure S5 of the ESI displays for further comparison the band structures of the five nearest neighbors to Si. Such examples demonstrate how our embeddings effectively captures both chemical and electronic similarities between materials, potentially accelerating the discovery of functional analogues to known materials.

Second, using the conventional superconductor \ce{MgB2} as a case study, we identified structurally and chemically related compounds with potentially similar properties. The closest compound, \ce{Mg4AlB10} (distance 0.16), is a five-layer supercell of \ce{MgB2} with one Mg layer replaced by Al, preserving similar band dispersion patterns. Interestingly, our method identified \ce{LiMgB4} (which lies 0.123~eV/atom above the convex hull) as having a small distance to \ce{MgB2}, with a calculated superconducting critical temperature of approximately 20.8~K. The band structures of the five most similar systems to \ce{MgB2} can be examined in the ESI. This example illustrates how our embeddings can uncover promising candidates with a desired functionality without performing any calculation, as the proximity in our representation space correlates with comparable physical properties.

Third, we introduce a new descriptor for materials on the convex hull: the ``uniqueness'' U, defined as the minimal Euclidean distance to neighboring materials in our embedding space. A high value of this metric reveals truly distinctive compounds that may merit special attention in experimental studies. We found that As is the most unique elementary substance (distance 1.06 to P), while hexagonal BN is the most unique binary compound on the convex hull (distance 1.17 to graphite). Despite being isovalent and isostructural, hexagonal BN and graphite exhibit dramatically different electronic properties: the former is a large-gap insulator while the latter is a semimetal. This measure of uniqueness offers valuable insights for both theoretical and experimental materials design by highlighting compounds with few similar alternatives.

In conclusion, we propose a fingerprint that provides a simple and interpretable representation of a compound, considering both its chemical composition and crystal structure. This is designed so that the distance between embeddings of compounds that are chemically similar --- and therefore likely to have similar materials properties --- is smaller than for unrelated materials. We show how this fingerprint can be used to create structure maps, or more generally property maps, spanning entire families of materials, or even the entire material space. Such maps can be used to visualise and interpret how materials properties change across chemical space. Finally, we show that our human-generated fingerprint can compete with machine-learned opaque representations when it is used as input feature to machine learning models. There are still some shortcomings in our approach, such as the lack of information for rare gases or some actinides, but we believe that our representation of the chemical space can already be an important tool both for accurate machine prediction of material properties and for human interpretation of high-throughput investigations.

\section{Methods}
    \label{sec:method}

\begin{figure}[tb]
    \centering
    \includegraphics[width=4cm]{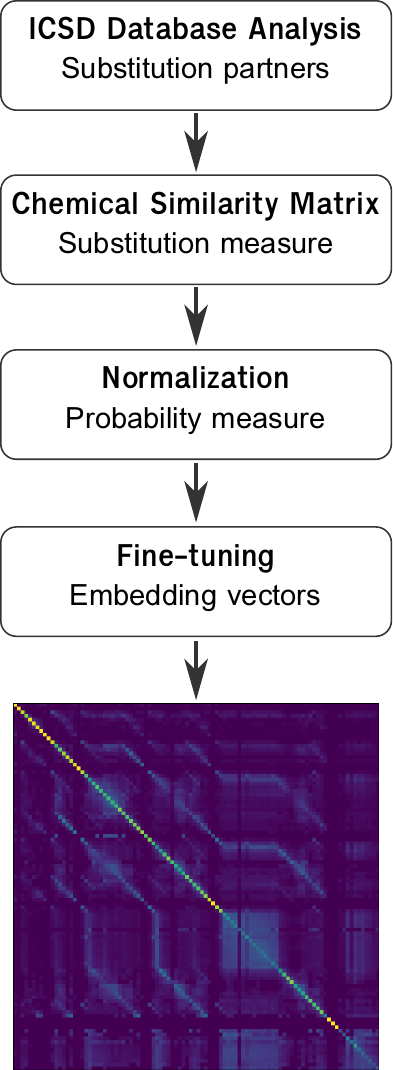}

    \caption{Schema depicting the construction of the matrix $\cal S$. Each line of the matrix is an unit vector that represents a chemical elements in the non-orthogonal compositional space.}
    \label{fig:schema}
\end{figure}

We start with the raw data from Ref.~\cite{wang2021predicting}, which is in the form of a (symmetric) matrix whose dimensions are given by the total number of chemical elements. The off-diagonal elements of the matrix count, for each pair of chemical elements (A, B), the number of compounds in ICSD that have the same crystal structure but where A is replaced by B. In the diagonal elements of the matrix we insert the total number of compounds in ICSD that contain the given element (which can be interpreted as self-substitutions). We then normalise the rows to one so that the entries can be interpreted as a measure of similarity.

Mainly due to the incomplete information present in the ICSD, the off-diagonal components are underestimated with respect to the diagonal. To compensate for this, we decided to modify the matrix elements by raising them by a power of $\alpha=1/2$, followed by a renormalisation of the lines. The resulting matrix is called $\cal S$. We then interpret each row $\mathcal{S}_i$ of the matrix $\cal S$ as the Cartesian coordinates of the unit vector representing the corresponding chemical element. This means that the off-diagonal components are the cosines of the angles formed by the unit vectors defining a non-orthogonal compositional space. Completely dissimilar chemical elements are represented by orthogonal unit vectors, with the angle decreasing as the similarity increases. Note that while the non-orthogonal space still has a dimension equal to the number of chemical elements (since all elements are dissimilar to some extent), the hypercube generated by the non-orthogonal vectors has a smaller volume than in Cartesian space. A schema describing this construction can be found in \cref{fig:schema}.

\begin{figure}[tb]
    \centering
    \includegraphics[width=6cm]{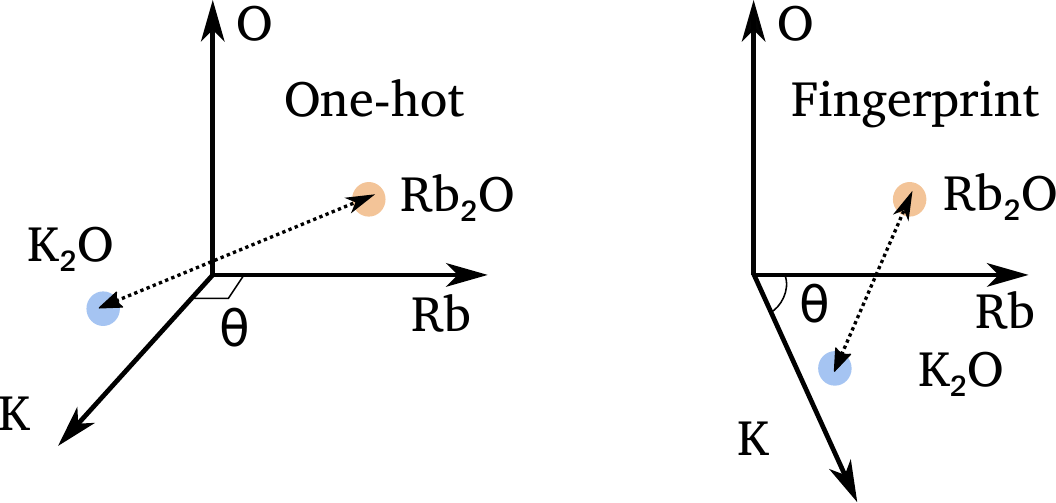}

    \caption{Schema depicting the Euclidean distance between \ce{K2O} (blue dot) and \ce{Rb2O} (brown dot) in a one-hot representation and in our fingerprint space.}
    \label{fig:fingerprint}
\end{figure}

To obtain the fingerprint of a given composition, we first create a one-hot composition vector $c$, whose dimension is given by the total number of chemical elements and is then normalised so that it lies in the unit hyper-sphere in the non-orthogonal space. The fingerprint $f$ is then given in Cartesian coordinates by $f = c \times \cal S$. The distances between the chemical compositions, which measure their dissimilarity, are then simply calculated as the Cartesian distance between the fingerprints. 

An example is shown in \cref{fig:fingerprint} where we plot \ce{K2O} (blue dot) and \ce{Rb2O} (brown dot) in a standard one-hot representation and in our fingerprint space. The angle between K and Rb in our space is about 43$^\circ$ (while K and Rb remain orthogonal to O due to their dissimilarity), making the distance between the two compounds closer than in a one-hot Euclidean representation.

We call the composition embeddings Pettifor embeddings. By concatenating the latter with crystal structure embeddings~\cite{zimmermann2017assessing} we obtain the  Pettifor$\oplus$STR embeddings.

\section{Data and code availability}

The data used in this work, along with the accompanying example notebook, is available on GitHub at \url{https://github.com/hyllios/utils/tree/main/similarity}

\section{Acknowledgements}
T.F.T.C acknowledges financial support from Fundação para a Ciência e Tecnologia (FCT), I.P. through the project CEECINST/00152/2018/CP1570/CT0006 \newline{}with DOI identifier 10.54499/CEECINST/00152/2018/\newline{}CP1570/CT0006. S.B. acknowledges funding from the Volkswagen Stiftung (Momentum) through the project “dandelion” and from the Deutsche Forschungsgemeinschaft (DFG, German Research Foundation) through the project BO 4280/11-1.

\section{Author  Contributions}
T.F.T.C. and M.A.L.M. developed the Pettifor embedding method. T.F.T.C. and H.C.W. implemented the method in Python. T.F.T.C. trained the CrabNet models. T.F.T.C. and H.C.W. prepared the figures. All authors contributed to the editing and revision of the manuscript. M.A.L.M. and S.B. supervised the project and secured funding.

\section{Competing  Interests}
The authors declare that they have no competing interests.


%

\end{document}